\RequirePackage[T1]{fontenc}
\documentclass[journal]{IEEEtran}
\usepackage{cite}
\usepackage{amsmath,amssymb,amsfonts}
\usepackage{graphicx}
\usepackage{textcomp}
\usepackage{xcolor}
\usepackage{booktabs}
\usepackage{multirow}
\usepackage{array}
\usepackage{xurl}
\usepackage{bm}
\usepackage{balance}
\usepackage{algorithmic}
\usepackage{stfloats}
\usepackage{placeins}
\usepackage[bookmarks=false,hidelinks]{hyperref}

\usepackage{algorithm}
\usepackage{algorithmic}

\graphicspath{{figure_new/}}

\AtBeginDocument{%
\setlength{\abovedisplayskip}{5pt plus 1pt minus 1pt}
\setlength{\belowdisplayskip}{5pt plus 1pt minus 1pt}
\setlength{\abovedisplayshortskip}{4pt plus 1pt minus 1pt}
\setlength{\belowdisplayshortskip}{4pt plus 1pt minus 1pt}}

\newcommand{\CN}{\mathcal{CN}}

\newcommand{\diag}{\operatorname{diag}}

\newcommand{\Prb}{\operatorname{Pr}}

\begin{document}

\title{Vision--RF Cooperative ISAC for Dynamic V2I Networks: Adaptive Beam Management and Semantic Clutter Suppression}

\author{Wenqi~Fan, Ning~Wei, \emph{IEEE Member}, Xiangyu Dong,  Ahmad~Bazzi, \emph{IEEE Senior Member}, \\
Guan Gui, \emph{IEEE Fellow}, Rongyan~Xi, Huimin Tang, and Yue~Xiu, \emph{IEEE Member}\\
\thanks{Wenqi Fan, Ning Wei, Xiangyu Dong, and Yue Xiu are with the University of Electronic Science and Technology of China, Chengdu, China. (E-mail: fwq897032532@gmail.com, wn@uestc.edu.cn, dong\_xiangyu@yeah.net, xiuyue12345678@163.com). Ahmad Bazzi is with the Wireless Research Lab, New York University, New York 10012 United States (E-mail:ahmad.bazzi@nyu.edu). Guan Gui is with the College of Telecommunications and Information Engineering, Nanjing University of Posts and Telecommunications, Nanjing 210003, China (E-mail:guiguan@njupt.edu.cn).
Rongyan Xi is with China Mobile Research Institute, Beijing 10032 China(E-mail:xirongyan@chinamobile.com). Huimin Tang is with the School of Transportation and Logistics, Southwest Jiaotong University (E-mail: tanghuimin@my.swjtu.edu.cn).}}
\maketitle
\begin{abstract}
We develop a vision--radio-frequency (RF) cooperative integrated sensing and communication (ISAC) framework that jointly uses visual priors and RF observations for transmit-side beam control and receiver-side clutter suppression. Within a narrowband snapshot model, visual target priors and historical RF detections guide control in dynamic vehicle-to-infrastructure networks with roadside clutter. In this framework, posterior angular uncertainty determines beam control and the selection between communication-for-sensing and sensing-for-sensing modes. To reduce sensing overhead while accommodating unreliable visual information, we develop a reliability-adaptive beam scanning scheme that allocates a bounded scan budget according to visual target probabilities and RF history, while reserving an adaptive fraction for exploration. At the receiver, visual geometry identifies clutter support, and RF observations estimate the corresponding complex reflection coefficients. Motion-aware residual filtering is then applied before angle--Doppler detection to suppress clutter while preserving slow-moving targets. Furthermore, communication and sensing power allocation is formulated as a per-direction linear program, subject to signal-to-interference-plus-noise ratio constraints evaluated at angles sampled from the posterior distribution and a total-power budget. Simulation results show that the proposed framework improves tracking over static and vision-only beam control and reduces sensing overhead relative to full scanning. The proposed scanning scheme also provides greater robustness to visual blockage than fixed-weight scanning, while the receiver better preserves slow-moving targets than temporal cancellation when visual target information is available.
\end{abstract}


\section{Introduction}


Millimeter-wave (mmWave) integrated sensing and communication (ISAC) systems offer substantial bandwidth for vehicle-to-infrastructure (V2I) networks. Narrow beams are employed to compensate for propagation loss, but maintaining beam alignment becomes challenging as vehicles move \cite{heath2016,7905941,kalamkar2022beam}. In particular, a roadside mmWave ISAC base station (BS) can reuse its antenna array and transmitted waveforms to serve vehicles while sensing surrounding targets \cite{liu2022isac}. However, the coexistence of communication users, sensing targets, and static roadside objects complicates practical operation. Communication beams must be updated to maintain service for moving vehicles, while communication or dedicated sensing beams are used to illuminate other targets. Meanwhile, reflections from roadside objects introduce clutter that can impair target detection. These challenges motivate the joint consideration of three closely related tasks: 1) beam tracking for moving communication users, 2) beam scanning for sensing targets, and 3) clutter suppression at the sensing receiver.


In this paper, we develop a vision--radio-frequency (RF) framework for beam control, sensing beam scanning, and clutter suppression in roadside ISAC systems. Visual observations provide object locations, geometry, and semantic labels that complement RF measurements \cite{11098592,Xiu2026Robust,Guo2026Embodied,cheng2024synesthesia,camera_radar_review}. This information can guide beams toward likely users or targets before an exhaustive RF scan. 
However, occlusion, missed detections, and angular estimation errors
can degrade both beamforming gain and communication--sensing
interference control. We therefore combine visual observations, state
prediction, and RF updates within a common narrowband framework.

We first consider beam control under uncertain user angles. Instead of relying only on point estimates, we maintain a posterior distribution of user angles through state prediction and measurement updates, including periods of intermittent visibility. The posterior covariance is incorporated into beam control because angular uncertainty affects the desired communication gain, inter-user interference, and interference from dedicated sensing beams. This provides a common basis for beam protection and communication constraints evaluated at angles sampled from the posterior distribution. To reduce the overhead of full angular scanning, we further develop a reliability-adaptive sensing beam scanning scheme. Visual target probabilities and historical RF detections are combined to prioritize sensing directions within a bounded scan budget. Since fixed visual priorities may neglect occluded or missed targets, an adaptive fraction of the budget is reserved for exploration. We consider two sensing modes: communication-for-sensing (C4S), which reuses communication beams, and sensing-for-sensing (S4S), which introduces a dedicated sensing beam \cite{luo2024clutter_isac}. Mode selection and power allocation account for posterior angular uncertainty to balance target illumination against communication interference.


At the receiver, we consider the structured clutter produced by buildings, trees, parked vehicles, and other roadside objects. Conventional temporal averaging and moving-target-indication filtering suppress nearly static returns, but may also attenuate slow-moving targets whose Doppler phase changes little over a short observation interval \cite{richards2014}. To address this limitation, we use visual geometry and semantic information to identify candidate clutter support and vehicle regions, while estimating the corresponding complex reflection coefficients from RF observations. The reconstructed clutter is then suppressed before temporal filtering. Furthermore, motion information is used to moderate residual subtraction in regions likely to contain targets, thereby balancing clutter suppression against target preservation.
Finally, we evaluate the transmission and receiver designs while distinguishing angular coverage from signal-to-interference-plus-noise ratio (SINR) feasibility, and retained target power from detection performance. For clarity, the main contributions of this paper are summarized as follows:

\begin{itemize}

\item We combine covariance-aware vision--RF state estimation with reliability-adaptive sensing beam scanning (SBS). Posterior covariance determines beam steering and Gaussian angular protection for uncertainty-aware C4S/S4S selection; visual reliability and RF history determine scan priorities with adaptive exploration and a finite revisit bound.



\item We formulate per-direction power allocation problems under SINR constraints evaluated at angles sampled from the posterior distribution, and derive a spectral-radius condition for feasibility.

\item We develop a clutter suppression method that combines maximum a posteriori (MAP) estimation of clutter coefficients over visually identified support with motion-aware residual filtering. We further derive target-leakage and retained-power properties, with class-conditioned variances incorporated as optional priors.

\end{itemize}
Section~II reviews related work. Sections~III--V present the system model, transmitter, and receiver; Sections~VI and~VII provide simulation results and conclusions.

\section{Related Work}

\subsection{Multimodal and Vision-Assisted ISAC}


Visual information has been considered as a source of side information for wireless communications, where image and depth observations provide environmental knowledge that would otherwise require additional channel measurements. Existing studies have investigated its use for beam selection, received-power prediction, and blockage prediction \cite{Xiu2025MovableAirborne,Lyu2025MovableISAC,10870062,alrabeiah2020vision,nishio2020vision,charan2021vision}. To further exploit complementary modalities, cross-modal feature enhancement and dynamic fusion have been introduced for beam prediction \cite{zhu2025multimodalbeam}. In addition, multimodal ISAC extends the use of environmental information to both communication and sensing, while camera--radar fusion combines visual semantics with RF observations for perception \cite{cheng2024synesthesia,camera_radar_review}. A predicted beam or learned feature representation does not, by itself, provide an explicit angular confidence interval for subsequent power allocation. Such uncertainty information is needed because angular errors affect both the desired communication gain and interference from other beams. To account for this uncertainty, we employ Gaussian state prediction and information-form fusion \cite{sarkka2013} to explicitly obtain the posterior mean and covariance. The posterior mean guides beam steering, while the covariance determines angular protection regions and quality-of-service (QoS) constraints evaluated at sampled posterior angles. Unlike beam-index predictions or point estimates alone, this formulation propagates posterior covariance into beam protection, C4S/S4S selection, and sampled-angle QoS constraints. State prediction and available RF updates maintain this representation during visual unavailability.

\subsection{Dynamic Beam Management and Adaptive Sensing}


Beam management has been investigated to maintain alignment under user mobility while reducing measurement overhead \cite{7905941,kalamkar2022beam}. Related studies on sensor scheduling and cognitive radar allocate sensing resources according to tracking requirements, state uncertainty, and information measures \cite{sensor_scheduling_tracking}. In ISAC systems, C4S and S4S provide two options for target illumination, using communication beams and dedicated sensing beams, respectively \cite{luo2024clutter_isac}. In addition, robust optimization and interference-coupled power control provide methods for resource allocation under uncertainty and mutual interference \cite{bental2009,yates1995}. Visual guidance may overlook targets that are absent from the camera map due to occlusion or missed detections. To address this limitation, we adapt the contributions of visual information and RF history according to their observed reliability, while reserving part of the scan budget for exploration. Directions are revisited in an oldest-first order to ensure a finite revisit bound. Furthermore, posterior angular covariance guides C4S/S4S mode selection, and communication and sensing power are allocated under interference constraints evaluated at sampled posterior angles. The resulting feasibility condition applies to these sampled angular scenarios; it does not establish feasibility over the entire continuous uncertainty region. Relative to fixed directional priorities or predetermined sensing sectors, our scheduler updates priorities using visual reliability, RF history, and angular uncertainty while retaining exploration and revisits.

\subsection{Clutter Modeling and Slow-Target Protection}


Static-reflector models have been used to characterize environmental returns, while temporal averaging and moving-target-indication (MTI) filtering suppress components concentrated near zero Doppler \cite{luo2024clutter_isac,richards2014}. In addition, constant-false-alarm-rate (CFAR) processing determines detection thresholds from neighboring background estimates, which can be affected by nonhomogeneous clutter and nearby targets \cite{gandhi1988}. Clutter cancellation and adaptive thresholding therefore operate at different stages of the detection process. Over a finite observation interval, slow-moving targets may exhibit temporal characteristics similar to those of static reflectors. Consequently, stronger temporal subtraction may also remove target energy before CFAR processing is applied.
To address this limitation, we use scene semantics to identify likely clutter support and distinguish it from regions containing moving objects. However, visual information does not directly determine the amplitudes and phases of RF reflections. Visual semantics and geometry determine the spatial support and classes of candidate clutter components; RF snapshots estimate their complex coefficients using MAP estimation \cite{kay1993}. The estimated RF amplitudes and phases are associated with the visually selected support to reconstruct and subtract the complex clutter field from received snapshots. Following clutter reconstruction and cancellation, motion-aware residual filtering moderates temporal subtraction in vehicle regions to limit the attenuation of slow-moving targets. Our analysis distinguishes reconstruction errors and target leakage in MAP estimation from the attenuation introduced by subsequent temporal filtering. Furthermore, clutter suppression, retained target power, and detection performance are evaluated separately, since greater target-power retention does not necessarily improve detection in the presence of residual clutter. Compared with cancellation based on temporal RF structure alone, our receiver combines visual support selection, RF coefficient calibration, and motion-aware slow-target protection.

\section{System Model}

\begin{figure}[t]
\centering
\includegraphics[width=\columnwidth
]{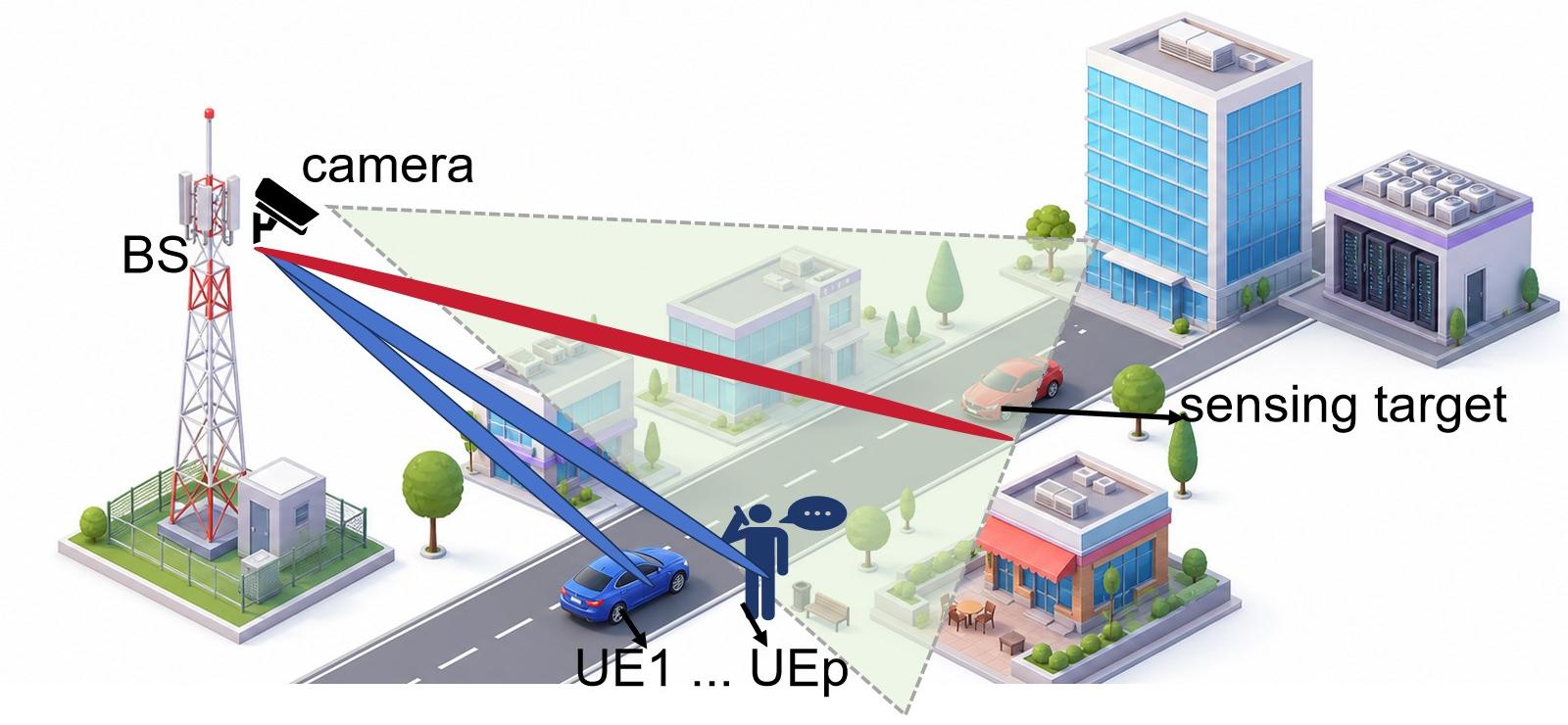}
\vspace{-20pt}
\caption{Roadside vision--RF ISAC with dynamic users, sensing targets, and environmental clutter.}
\label{fig:system_scenario}
\end{figure}
\subsection{Roadside V2I Network and Arrays}

We consider a time-slotted roadside BS serving $P$ dynamic user equipments (UEs) and sensing targets in environmental clutter, as shown in Fig.~\ref{fig:system_scenario}. Blue and red beams indicate communication and sensing, respectively. The BS is equipped with $N_t$ transmit and $N_r$ receive uniform linear array (ULA) elements, a narrowband RF front end, and a calibrated depth/stereo camera. Slot $n$ has duration $T_{\rm slot}$, with $Q_s$ matched-filtered snapshots separated by $T_p$ per sensing direction. RF sensing resolves angle and Doppler; range for association comes from vision or external coarse ranging, since single-carrier phase does not resolve absolute range.
For concreteness, we instantiate the framework using ULAs. The control and fusion framework extends to other calibrated array geometries by replacing the array-response and coupling functions with their corresponding forms. According to the standard ULA model \cite{balanis2016}, for an $N$-element array with spacing $d_\lambda$ measured in wavelengths, the array response vector is defined as
\begin{equation}
\mathbf a_N(\theta)=
\left[1,e^{j2\pi d_\lambda\sin\theta},\ldots,
e^{j2\pi d_\lambda(N-1)\sin\theta}\right]^T,
\end{equation}
and the normalized power coupling
\begin{equation}
G_N(\theta,\vartheta)=
|\mathbf a_N^H(\theta)\mathbf a_N(\vartheta)|^2/{N^2}.
\end{equation}
Hence, $0\leq G_N\leq1$. The full half-power beamwidth around $\vartheta$ is approximated by the standard large-array expression \cite{balanis2016}
\begin{equation}
b_{\rm 3dB}(\vartheta)=
{0.88}/{N_t d_\lambda\cos\vartheta},
\end{equation}
with clipping to $\mathcal A=[-\pi/2+\delta_\theta,\pi/2-\delta_\theta]$, where $\delta_\theta>0$ excludes the endfire singularities of the beamwidth approximation. This interval models the visible front half-space of the ULA; the scan codebook may cover a narrower field.

\subsection{Dynamic User State and Multimodal Observations}

The range, angle, and their rates form vehicle $p$'s state $\boldsymbol\xi_p(n)=[r_p(n),\theta_p(n),\dot r_p(n),\dot\theta_p(n)]^T$.
The constant-velocity (CV) model is given by \cite{li2003survey}
\begin{equation}
\boldsymbol\xi_p(n+1)=\mathbf F\boldsymbol\xi_p(n)+\mathbf w_p(n),
\quad \mathbf w_p(n)\sim\mathcal N(\mathbf 0,\mathbf Q),
\end{equation}
where
\begin{equation}
\mathbf F=
\begin{bmatrix}
1&0&T_{\rm slot}&0\\
0&1&0&T_{\rm slot}\\
0&0&1&0\\
0&0&0&1
\end{bmatrix}.
\end{equation}
The process-noise covariance $\mathbf Q$ characterizes the uncertainty caused by unmodeled radial and angular acceleration, including the resulting uncertainty in range, angle, radial velocity, and angular
velocity during state prediction. The Kalman prediction step
\cite{sarkka2013} produces

\begin{equation}
\widehat{\boldsymbol\xi}_p^- =\mathbf F\widehat{\boldsymbol\xi}_p,
\qquad
\boldsymbol\Sigma_p^- =\mathbf F\boldsymbol\Sigma_p\mathbf F^T+\mathbf Q.
\label{eq:state_prediction}
\end{equation}
When vehicle $p$ is visible to the camera, the visual processing module provides estimates of its range, bearing angle, and angular velocity from calibrated image geometry, 3-D object detection, and
temporal tracking \cite{xu2023visionbeam,dong2022depth}. We model these estimates as a Gaussian observation,
\begin{equation}
\mathbf z_p^v=\mathbf H_v\boldsymbol\xi_p+\mathbf v_p,\qquad
\mathbf H_v=
\begin{bmatrix}
1&0&0&0\\0&1&0&0\\0&0&0&1
\end{bmatrix},
\end{equation}
where $\mathbf v_p\sim\mathcal N(\mathbf0,\mathbf R_p^v)$ \cite{5be8c321f17d4189a3ed780eefc7b185}. Occlusion reduces visual detection probability, and lower confidence increases $\mathbf R_p^v$. The camera also reports target vehicle class and moving probability. The RF tracker measures angle and radial velocity as
\begin{equation}
\mathbf z_p^r=\mathbf H_r\boldsymbol\xi_p+\mathbf e_p,\qquad
\mathbf H_r=
\begin{bmatrix}0&1&0&0\\0&0&1&0\end{bmatrix},
\end{equation}
where $\mathbf e_p\sim\mathcal N(\mathbf0,\mathbf R_p^r)$ and the covariance decreases with matched-filter signal-to-noise ratio (SNR). For modality $m\in\{v,r\}$, $\mathbf H_p^m$, $\mathbf R_p^m$, and $\mathbf z_p^m$ denote its observation matrix, measurement-noise covariance, and measurement, respectively; $\mathbf H_p^v=\mathbf H_v$ and $\mathbf H_p^r=\mathbf H_r$. For the available, conditionally independent Gaussian observations, information-form fusion \cite{sarkka2013,kay1993} gives
\begin{align}
\boldsymbol\Sigma_p(n)=\left[(\boldsymbol\Sigma_p^-)^{-1}+\sum_{m\in\mathcal O_p(n)}(\mathbf H_p^m)^T
(\mathbf R_p^m)^{-1}\mathbf H_p^m\right]^{-1},
\label{eq:fusion_covariance}\\
\widehat{\boldsymbol\xi}_p(n)\!=\!\boldsymbol\Sigma_p(n)\!\left[
(\boldsymbol\Sigma_p^-)^{-1}\widehat{\boldsymbol\xi}_p^-
+\sum_{m\in\mathcal O_p(n)}(\mathbf H_p^m)^T
(\mathbf R_p^m)^{-1}\mathbf z_p^m\right],
\label{eq:fusion_mean}
\end{align}
where $\mathcal O_p(n)\subseteq\{v,r\}$ contains the available modalities. We denote the fused angle and variance by $\widehat\theta_p(n)$ and $\sigma_{\theta,p}^2(n)=[\boldsymbol\Sigma_p(n)]_{2,2}$.

\subsection{Narrowband Communication and Sensing Signal Model}

The transmit signal at snapshot $q\in\{0,\ldots,Q_s-1\}$ for sensing direction \(d\) in slot \(n\) is given by
\begin{align}
\mathbf x_{d,q}(n)=
&\sum_{p=1}^{P}\sqrt{\frac{\rho^c_{p,d}(n)P_t}{N_t}}
\mathbf a_{N_t}(\widehat\theta_p(n))s^c_{p,d,q}(n)\nonumber\\
&+\chi_d(n)\sqrt{\frac{\rho^s_d(n)P_t}{N_t}}
\mathbf a_{N_t}(\Theta_d(n))s^s_{d,q}(n),
\label{eq:narrowband_transmit_signal}
\end{align}
Here $s^c_{p,d,q}(n)$ denotes user $p$\textquotesingle s communication/pilot symbol in sensing burst $d$ at snapshot $q$, and $s^s_{d,q}(n)$ denotes the dedicated sensing symbol. Symbols have unit average energy, and communication symbols are mutually uncorrelated for the SINR power accounting below. The equivalent sensing model in Section~V-A uses the known, mutually orthogonal unit-energy pilot sequences specified there. The total transmit-power budget is $P_t$, $\Theta_d(n)$ is the selected sensing direction, and $\chi_d(n)\in\{0,1\}$ indicates whether a dedicated sensing beam is active. All $P$ communication beams are maintained, with their steering directions determined by the fused angle estimates. The pilot symbols are known to the BS, and the nonnegative power allocation coefficients $\rho_{p,d}^c$ and $\rho_d^s$ specify the fractions of $P_t$ assigned to communication and dedicated sensing, respectively. In C4S mode, $\chi_d=0$ and $\rho_d^s=0$. In S4S mode, $\chi_d=1$, enabling a dedicated sensing beam that also introduces interference to the communication users. The downlink channel from the BS to vehicle $p$ is modeled as a single-path narrowband mmWave channel \cite{heath2016}, given by
\begin{equation}
\mathbf h_p(n)=\sqrt{\beta_p^c(n)}\,\mathbf a_{N_t}(\theta_p(n)),
\label{eq:communication_channel}
\end{equation}
where \(\beta_p^c(n)\) denotes the large-scale channel power gain, modeled as
\begin{equation}
\beta_p^c(n)=C_0\left(\frac{r_p(n)}{r_0}\right)^{-\alpha_{\rm pl}},
\label{eq:large_scale_gain}
\end{equation}
where $C_0$ denotes the channel power gain at the reference distance $r_0$, and $\alpha_{\rm pl}$ is the path-loss exponent. Shadow fading is neglected. The unit-norm beamforming vector for user $l$ is defined as $\mathbf w_l^c(n)=\mathbf a_{N_t}(\widehat\theta_l(n))/\sqrt{N_t}$. Accordingly, the power received by user $p$ from the $l$th communication beam is given by
\begin{equation}
P_{p,l}^c(n)=\rho_{l,d}^c(n)P_t\beta_p^c(n)N_t
G_{N_t}(\theta_p(n),\widehat\theta_l(n)).
\label{eq:received_beam_power}
\end{equation}
Here, the spatial coupling depends on the actual user angle $\theta_p$ and the beam steering angle $\widehat\theta_l$ obtained from sensor fusion. The communication channel accounts for one-way propagation, whereas the sensing reflection coefficients introduced below incorporate round-trip propagation. Assuming mutually uncorrelated communication symbols \cite{heath2016,yates1995}, the aggregate multi-user interference power at user $p$ is expressed as
\begin{equation}
I^c_{p,d}=\sum_{l\ne p}\rho^c_{l,d}P_tN_t\beta_p^c
G_{N_t}(\theta_p,\widehat\theta_l).
\label{eq:SINR}
\end{equation}
According to \eqref{eq:SINR}, the downlink SINR is expressed as
\begin{equation}
\Gamma_{p,d}=
\frac{\rho^c_{p,d}P_tN_t\beta_p^cG_{N_t}(\theta_p,\widehat\theta_p)}
{I^c_{p,d}+\chi_d\rho^s_dP_tN_t\beta_p^c
G_{N_t}(\theta_p,\Theta_d)+\sigma_c^2}.
\label{eq:communication_sinr}
\end{equation}
Here $\sigma_c^2$ denotes the communication-noise power. The SINR in \eqref{eq:communication_sinr} is evaluated using the true user angle, whereas resource allocation is based on a finite grid of angles derived from the posterior distribution and the fused range estimate. Over each short sensing burst, the geometry and reflection coefficients are assumed to remain constant. Under the narrowband monostatic array model \cite{richards2014,sturm2011}, the received sensing signal at snapshot \(q\) for sensing direction \(d\) is expressed as follows, with the slot index \(n\) omitted for notational simplicity:

\begin{equation}
\begin{aligned}
\mathbf y_{d,q}^{r}
&=\sum_{k=1}^{K_d}\alpha_k e^{j2\pi f_{D,k}qT_p}
\mathbf a_{N_r}(\theta_k)\mathbf a_{N_t}^H(\theta_k)\mathbf x_{d,q}\\
&\quad+\sum_{i=1}^{L_c}\beta_i\mathbf a_{N_r}(\theta_i^{\rm cl})
\mathbf a_{N_t}^H(\theta_i^{\rm cl})\mathbf x_{d,q}+\mathbf n_{d,q},
\end{aligned}
\label{eq:received_array_echo}
\end{equation}
Here $\theta_k$ and $\theta_i^{\rm cl}$ are target and clutter bearings, and $\lambda$ is the carrier wavelength. The quantities \(K_d\) and \(L_c\) denote the numbers of targets and clutter cells, respectively. The complex coefficients \(\alpha_k\) and \(\beta_i\) account for the round-trip attenuation and constant carrier phase associated with the unresolved propagation delays. The Doppler frequency of target \(k\) is \(f_{D,k}=2v_k/\lambda\), where \(v_k\) denotes the radial velocity defined as positive in the direction yielding a positive Doppler shift. Static clutter is assumed to have zero Doppler shift. The receiver noise is modeled as \(\mathbf n_{d,q}\sim\mathcal{CN}(\mathbf0,\sigma_s^2\mathbf I_{N_r})\).
Section~V-A expands the communication and sensing contributions to \eqref{eq:received_array_echo} before receiver processing.

\subsection{Visual Target and Semantic Clutter Representations}
The camera detections form $\mathcal V(n)=\{(\widehat\theta_j^v,c_j,
\sigma_{\theta,j}^v,\gamma_j^v,p_{j}^{\rm mov})\}$. Here $\widehat\theta_j^v$ and $\sigma_{\theta,j}^v$ are the visual bearing estimate and its standard deviation, $c_j$ is the object class, $\gamma_j^v\in[0,1]$ is detection confidence, and $p_j^{\rm mov}\in[0,1]$ is the current moving-object probability. The semantic clutter map at slot \(n\) is represented by
\begin{equation}
\begin{aligned}
\mathcal M_v(n)=\{(&\widehat\theta_i^v,c_i,\gamma_i^v,\mu_i^{\rm movable},\\
&p_i^{\rm static},p_i^{\rm motion},\widehat r_i^{\rm coarse})\}_{i=1}^{L_{\rm map}}.
\end{aligned}
\end{equation}
For candidate $i$, $\widehat\theta_i^v$, $c_i$, and $\gamma_i^v$ denote its visual bearing, semantic class, and detection confidence; $p_i^{\rm static}$ is its static probability and $\widehat r_i^{\rm coarse}$ is its visual or externally supplied coarse range. Here, \(L_{\mathrm{map}}\) denotes the number of visual candidates and need not equal the number of physical clutter cells \(L_c\). The parameter \(\mu_i^{\mathrm{movable}}\) characterizes class-level movability, whereas \(p_i^{\mathrm{motion}}\) represents the probability of current motion. For example, a parked vehicle may have high movability but a low probability of current motion. The candidate clutter support is selected as

\begin{equation}
\begin{aligned}
\mathcal I_c=\{i:\;&p_i^{\rm static}\geq\tau_{\rm static},
\ p_i^{\rm motion}\leq\tau_{\rm mov},\\
&d(i,\mathcal V_{\rm mov})>\Delta_{\rm mask}\}.
\end{aligned}
\label{eq:clutter_eligibility}
\end{equation}
Here $\tau_{\rm static}$ is the minimum static-probability threshold, $\tau_{\rm mov}$ is the maximum current-motion probability allowed for a static candidate, and $\mathcal V_{\rm mov}$ contains visual detections with $p_j^{\rm mov}>\tau_{\rm mov}$. The function $d(i,\mathcal V_{\rm mov})$ is the minimum angular separation from candidate $i$ to these moving-target region-of-interest (ROI) centers, with $d=+\infty$ for an empty set. The angular exclusion margin $\Delta_{\rm mask}$ protects these ROIs. Candidates within the masking margin \(\Delta_{\mathrm{mask}}\) are excluded regardless of their static classification. For each selected candidate, the unknown RF reflection coefficient may be assigned a class-conditioned prior, \(\beta_i\mid c_i\sim\mathcal{CN}(0,\nu_{c_i})\), where \(\nu_{c_i}\) is the corresponding prior variance. Thus, visual observations determine the candidate clutter support and class labels, while RF observations are used to estimate the reflection coefficients.

\section{Vision-Assisted Adaptive ISAC Transmission}

\begin{figure*}[!t]
\centering
\includegraphics[width=0.8\textwidth]{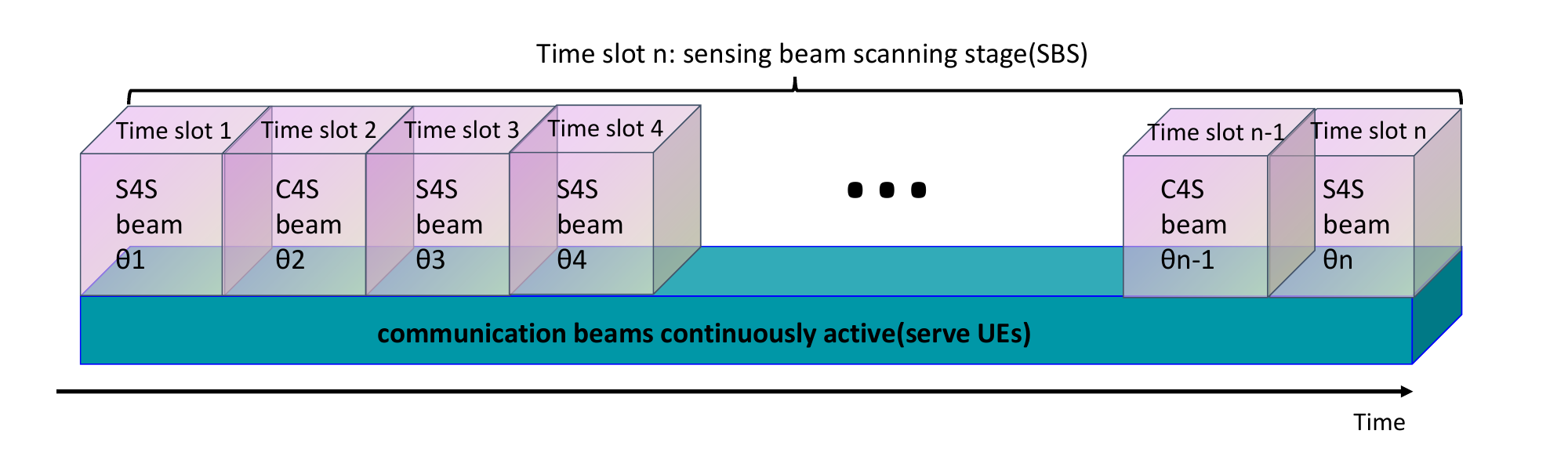}
\caption{Per-slot C4S/S4S beam scheduling with continuous
communication service during SBS.}
\label{fig:system_architecture}
\end{figure*}

Section~IV develops the transmit-side control stage of the proposed framework. Fig.~\ref{fig:system_architecture} illustrates the resulting per-slot
sensing-beam scheduling structure, where the selected directions are
sequentially operated in C4S or S4S mode while the communication
beams remain continuously active. Using the fused Vision--RF state and the semantic prior in Fig.~\ref{fig:overall_algorithm_flow}, this stage determines which directions should be sensed, whether each selected direction uses the C4S or S4S mode, and how the communication and sensing power should be allocated.


\subsection{Reliability-Adaptive Vision--RF SBS}

At slot \(n\), the scan priorities are determined using the current visual detections \(\mathcal V(n)\) and the RF detections from the preceding scan cycle, denoted by \(\mathcal H_r(n)\) \cite{sensor_scheduling_tracking}. The angular codebook contains $Q_{\rm full}$ directions $\{\Theta_g\}_{g=1}^{Q_{\rm full}}$. For notational simplicity, the slot index \(n\) is omitted from the following angular profiles. The visual target profile and the RF history score are defined as

\begin{equation}\label{eq:sbs_modality_priors}
\begin{aligned}
p_{\rm tar}^v(\Theta_g)&=
\frac{\sum_{j\in\mathcal V}\gamma_j^v
\mathcal N(\Theta_g;\widehat\theta_j^v,(\sigma_{\theta,j}^v)^2)}
{\sum_{g'}\sum_{j\in\mathcal V}\gamma_j^v
\mathcal N(\Theta_{g'};\widehat\theta_j^v,(\sigma_{\theta,j}^v)^2)},\\
s_r(\Theta_g)&=\sum_{\ell\in\mathcal H_r}w_\ell
\exp\!\left[-\frac{(\Theta_g-\widehat\theta_\ell^r)^2}{2\sigma_h^2}\right].
\end{aligned}
\end{equation}
Here $\Theta_g$ is codebook direction $g$, $\widehat\theta_\ell^r$ is historical RF bearing $\ell$, and $w_\ell=\gamma_\ell^r\delta^{a_\ell}$ discounts RF confidence by its age $a_\ell\geq1$ in scan cycles. The decay factor is $\delta\in(0,1)$, $\sigma_h$ is the RF-history kernel width, and $\gamma_\ell^r=1$ for accepted CFAR peaks; $\bar p_{\rm tar}^r=s_r/\max_g s_r$, or zero for empty history. The visual profile is zero for empty $\mathcal V$. The uncertainty profile $\bar u_{\rm unc}$ uses $\gamma_j^v\sigma_{\theta,j}^v$-weighted Gaussians of width $2\sigma_{\theta,j}^v$; the exploration profile $u_{\rm exp}$ favors directions not recently scanned. Bars denote unit-peak scaling.
Let $b_\ell=1$ if some visual angle lies within $2\sqrt{(\sigma_{\theta,j}^v)^2+\sigma_h^2}$ of RF detection $\ell$, and zero otherwise. The weighted agreement ratio is $c_R=\sum_\ell w_\ell b_\ell/\sum_\ell w_\ell$, with $c_R=1$ for empty RF history. Set $h_r=\mathbf1\{\mathcal H_r\ne\emptyset\}$. The parameter $\sigma_{\rm ref}$ is a reference visual angular uncertainty; $\lambda_{v0},\lambda_{r0},\lambda_{u0},\lambda_{e0}$ are baseline modality weights and $\kappa_e$ controls the exploration-weight increase. The reliability and utility are
\begin{equation}\label{eq:scan_utility}
\begin{aligned}
&r_v(n)=\frac{c_R}{|\mathcal V|}\sum_{j\in\mathcal V}\gamma_j^v
\exp\!\left[-\left(\frac{\sigma_{\theta,j}^v}{\sigma_{\rm ref}}\right)^2\right],\\
&\widetilde\lambda_v=\lambda_{v0}r_v,\quad
\widetilde\lambda_r=\lambda_{r0}h_r(1-r_v),\\
&\widetilde\lambda_e=\lambda_{e0}+[\kappa_e+\lambda_{r0}(1-h_r)](1-r_v),
\\
&U_{\rm scan}(\Theta_g,n)=\lambda_v\bar p_{\rm tar}^v+\lambda_r\bar p_{\rm tar}^r
 +\lambda_u\bar u_{\rm unc}+\lambda_e u_{\rm exp}.
\end{aligned}
\end{equation}
When \(\mathcal V=\varnothing\), the visual reliability is set to \(r_v=0\). The unnormalized uncertainty weight is fixed at \(\widetilde{\lambda}_u=\lambda_{u0}\), and the weights are normalized as
$\lambda_i=\widetilde\lambda_i/\sum_k\widetilde\lambda_k$, $i\in\{v,r,u,e\}$.
As $r_v$ decreases, more weight is assigned to RF history and
exploration; if RF history is unavailable, its weight is reassigned
to exploration. Neither the target count nor blockage probability
is required.
For \(\mathcal V\ne\varnothing\), let \(p_{(1)}\ge\cdots\ge p_{(Q_{\mathrm{full}})}\) denote the visual probability masses sorted in descending order. The coverage and uncertainty contributions to the scan budget are defined as
\begin{equation}
\begin{aligned}
&Q_{\rm cov}(n)=\min\left\{q:\sum_{r=1}^q p_{(r)}\geq\tau_{\rm cov}\right\},\\
&Q_{\rm unc}(n)=\left\lceil\kappa_Q\frac{\overline\sigma_\theta^v(n)}{\Delta_\Theta}\right\rceil,
\end{aligned}\label{eq:adaptive_budget_components}
\end{equation}
where $\tau_{\rm cov}\in(0,1)$ is the cumulative coverage threshold for target-directed visual probability, $\kappa_Q$ scales the uncertainty allowance, and \(\overline\sigma_\theta^v(n)\) is the confidence-weighted angular standard deviation and \(\Delta_\Theta\) is the angular grid spacing. The preliminary scan budget is $B_n=Q_{\rm cov}+Q_{\rm unc}+Q_{\rm exp,min}$.
\begin{equation}\label{eq:adaptive_scan_budget}
\begin{aligned}
&Q_n=\operatorname{clip}(B_n,Q_{\min},Q_{\max}),\\
&Q_{\rm exp}(n)=\min\{Q_n,q_1,q_0+\lceil\kappa_{\rm exp}(1-r_v)\rceil\},\\
&Q_{\rm tar}(n)=Q_n-Q_{\rm exp}(n),
\end{aligned}
\end{equation}
Here $Q_{\min}$ and $Q_{\max}$ bound the scan budget; $q_0=Q_{\rm exp,min}$ and $q_1=Q_{\rm exp,max}$ bound the exploration reserve. The parameter $\kappa_{\rm exp}$ controls its sensitivity to visual unreliability. For empty $\mathcal V$, we set $Q_n=Q_{\max}$ as a robustness fallback. Without visual target-directed support, a smaller budget could leave occluded or missed targets undiscovered; RF history and exploration therefore use the maximum allowed budget. This is a design choice, not an optimality result. 
Exploration and target-directed scanning share the total scan budget \(Q_n\); thus, allocating more directions to exploration reduces the budget available for target-directed scanning. The scheduler first selects the \(Q_{\mathrm{tar}}\) directions with the highest utility, followed by the \(Q_{\mathrm{exp}}\) least recently scanned directions from the remaining candidates. Let \(s_g(n)\in\{0,1\}\) indicate whether direction \(g\) is selected in slot \(n\). Let $A_g(n)$ count slots since direction $g$ was last scanned. Its update, total scanning time, and energy consumption are given by
\begin{equation}
\begin{aligned}
&A_g(n+1)=(1-s_g(n))[A_g(n)+1],\\
&T_{\rm SBS}(n)=Q_nQ_sT_p,\quad E_{\rm SBS}(n)=P_{\rm burst}Q_nQ_sT_p,
\end{aligned}\label{eq:scan_age_recursion}
\end{equation}
where $P_{\rm burst}$ denotes the average transmit power used for the
scan-energy calculation. The angular codebook contains $Q_{\rm full}$ directions; numerical parameters and their calibration procedure are reported in Section~VI and Table~\ref{tab:simulation_parameters}.

\subsection{Uncertainty-Aware Dynamic C4S/S4S}

Let \(\mathcal Z_n\) denote the observations available up to and including slot \(n\). The posterior distribution of the user angle is modeled as \(\theta_p\mid\mathcal Z_n\sim\mathcal N(\widehat\theta_p,\sigma_{\theta,p}^{2})\). For \(\sigma_{\theta,p}>0\), the normalized angular error \(e_p^\theta=(\theta_p-\widehat\theta_p)/\sigma_{\theta,p}\) satisfies \(e_p^\theta\mid\mathcal Z_n\sim\mathcal N(0,1)\). Accordingly, for any \(\kappa>0\), the posterior probability that the angular error lies within \(\kappa\) standard deviations is given by \cite{sarkka2013}
\begin{equation}
\begin{aligned}
&\Prb\{|\theta_p-\widehat\theta_p|\leq\kappa\sigma_{\theta,p}
\mid\mathcal Z_n\}
=\Prb\{-\kappa\leq e_p^\theta\leq\kappa\mid\mathcal Z_n\}\\
&\quad=\Phi(\kappa)-\Phi(-\kappa)=2\Phi(\kappa)-1,
\end{aligned}
\label{eq:c4s_gaussian_coverage}
\end{equation}
where $\Phi(\cdot)$ denotes the cumulative distribution function (CDF) of a standard Gaussian random variable. 
For \(\kappa=2\), (25) yields a posterior angular coverage probability of approximately \(95.45\%\). To account for angular uncertainty, the nominal half-power sector is extended by \(\kappa\sigma_{\theta,p}(n)\) on each side, yielding the dynamic protection width
\begin{equation}
B_p(n)=b_{\rm 3dB}(\widehat\theta_p(n))
+2\kappa\sigma_{\theta,p}(n).
\label{eq:dynamic_protection_width}
\end{equation}
\begin{figure}[t]
    \centering
    \includegraphics[width=0.98\columnwidth]
    {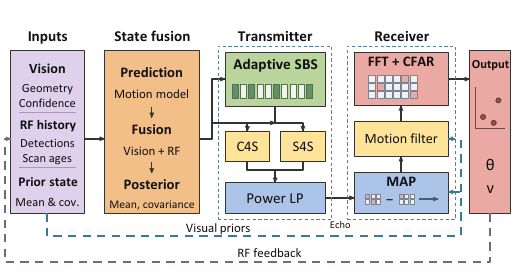}
    \caption{Per-slot processing flow of the proposed vision--RF
    environment-aware ISAC framework, illustrating the connection
    between the transmit-side control in Section~IV and the
    receiver-side processing in Section~V.}
    \label{fig:overall_algorithm_flow}
\end{figure}

This protection region is used solely for geometric guarding; it neither modifies the transmitted beam pattern nor guarantees the communication SINR. The probability in \eqref{eq:c4s_gaussian_coverage} refers to the untruncated Gaussian interval. When the posterior variance is zero, the protection width reduces to the nominal beamwidth, subject to the same cap. For a candidate sensing direction \(\Theta_d\), the normalized interference metric and its conditional exceedance probability are defined as
\begin{equation}
\begin{aligned}
Z_{p,d}(\rho_d^s)&=\rho_d^sG_{N_t}(\theta_p,\Theta_d),\\
\mathcal R_{p,d}(\rho_d^s)&=\Prb\{Z_{p,d}(\rho_d^s)>I_{\rm th}\mid\mathcal Z_n\},
\end{aligned}
\label{eq:interference_risk}
\end{equation}
where \(I_{\mathrm{th}}\) denotes the threshold for the normalized interference metric, and the probability is taken with respect to the posterior distribution of \(\theta_p\). Then, we evaluate \eqref{eq:interference_risk} by Gauss--Hermite quadrature. Direction $d$ is C4S-eligible for user $p$ when both the geometric protection test and the minimum sensing-gain test hold:
\begin{equation}
\begin{aligned}
|\Theta_d-\widehat\theta_p|&\leq B_p/2,\quad
G_{N_t}(\Theta_d,\widehat\theta_p)\geq G_{\rm C4S,min}.
\end{aligned}
\label{eq:protection_hit}
\end{equation}
C4S is included in the set of admissible modes if at least one user \(p\) satisfies both conditions in \eqref{eq:protection_hit}, whereas S4S is always included for subsequent feasibility evaluation. The threshold $G_{\rm C4S,min}$ specifies the minimum normalized sensing gain for C4S eligibility. For S4S, the maximum sensing-power fraction permitted by the interference-risk constraint is defined as
\begin{equation}
\overline\rho_d^s=\sup\left\{\rho\in[0,1]:
\max_p\mathcal R_{p,d}(\rho)\leq\epsilon_{\rm int}\right\}.
\label{eq:s4s_risk_cap}
\end{equation}
Here $\epsilon_{\rm int}\in(0,1)$ is the maximum allowed posterior probability of exceeding $I_{\rm th}$. The BS solves the power problem below for each admissible mode, enforces
$\rho_d^s\leq\overline\rho_d^s$ for S4S, and selects the feasible mode
with the larger sensing objective.
The interference risk is evaluated at the allocated sensing power. If neither mode is feasible, the BS applies the common fallback policy. The parameters \(\kappa\) and \(\kappa_c\) govern the geometric protection region and the robust QoS constraints, respectively.

\subsection{Robust Per-Direction Power Allocation}
\label{sec:power_allocation}

We formulate the sampled-envelope power allocation problem using a finite-scenario robust optimization framework \cite{bental2009}, in which the SINR requirements are expressed as affine constraints on the power allocation variables \cite{yates1995}. For user \(p\), the posterior angular credible interval is defined as
\begin{equation}
\mathcal U_p(n)=\left[\widehat\theta_p(n)-\kappa_c\sigma_{\theta,p}(n),
\widehat\theta_p(n)+\kappa_c\sigma_{\theta,p}(n)\right]
\label{eq:qos_credible_interval}
\end{equation}
Here $\kappa_c\geq0$ specifies the credible-interval half-width in posterior standard deviations. For an integer $S_c\geq2$, the interval is discretized into $S_c$ deterministic scenarios using normalized, equally spaced offsets $t_s\in[-1,1]$:
\begin{equation}
\begin{aligned}
\theta_{p,s}&=\operatorname{clip}_{\mathcal A}\!\left(
\widehat\theta_p+\kappa_c\sigma_{\theta,p}t_s\right),\\
t_s&=-1+\frac{2(s-1)}{S_c-1},\qquad s=1,\ldots,S_c,
\end{aligned}
\label{eq:qos_angle_scenarios}
\end{equation}
where $\operatorname{clip}_{\mathcal A}$ clips angles to the visible interval defined in Section~III-A. Odd $S_c$ includes the posterior mean; nominal control
($\kappa_c=0$) uses that mean alone. 
For scenario \(s\), the received-power coefficients per unit transmit-power fraction are given by
\begin{equation}
\begin{aligned}
a_{p,l}^{(s)}&=P_tN_t\widehat\beta_p^c
G_{N_t}(\theta_{p,s},\widehat\theta_l),\\
a_{p,d}^{s,(s)}&=P_tN_t\widehat\beta_p^c
G_{N_t}(\theta_{p,s},\Theta_d),
\end{aligned}
\label{eq:sampled_qos_coefficients}
\end{equation}
where $\widehat\beta_p^c$ follows \eqref{eq:large_scale_gain} with the fused
range. Holding the steering angles, direction, and mode fixed gives
\begin{equation}
\Gamma_{p,d}^{(s)}=
\frac{a_{p,p}^{(s)}\rho_{p,d}^c}
{\sum_{l\ne p}a_{p,l}^{(s)}\rho_{l,d}^c
+\chi_d a_{p,d}^{s,(s)}\rho_d^s+\sigma_c^2}.
\label{eq:scenario_sinr}
\end{equation}
Here $\sigma_c^2>0$ is the communication-noise power and $\Gamma_{\min}>0$ is the minimum required downlink SINR. Since $\sigma_c^2>0$, 
the denominator in \eqref{eq:scenario_sinr} is strictly positive. Hence, imposing \(\Gamma_{p,d}^{(s)}\ge\Gamma_{\min}\) for every sampled scenario is equivalent to the affine constraints
\begin{equation}
\begin{aligned}
&a_{p,p}^{(s)}\rho_{p,d}^c
-\Gamma_{\min}\sum_{l\ne p}a_{p,l}^{(s)}\rho_{l,d}^c
-\Gamma_{\min}\chi_d a_{p,d}^{s,(s)}\rho_d^s
\\&\geq\Gamma_{\min}\sigma_c^2,\quad\forall p,s.
\end{aligned}
\label{eq:scenario_sinr_constraints}
\end{equation}
Since \eqref{eq:scenario_sinr_constraints} is enforced for each of the $P$ users over all $S_c$ sampled angular scenarios, it gives a total of $P S_c$ affine SINR constraints. Together with the linear sensing objective and the power
constraints introduced below, these constraints define a finite-scenario linear program (LP). The evaluated controller uses the more conservative sampled envelopes
\begin{equation}
\begin{aligned}
\underline a_{p,p}&=\min_s a_{p,p}^{(s)},\\
\overline a_{p,l}&=\max_s a_{p,l}^{(s)},~l\ne p,\qquad
\overline a_{p,d}^{s}=\max_s a_{p,d}^{s,(s)}.
\end{aligned}
\label{eq:robust_qos_envelopes}
\end{equation}
Since the power allocation variables are nonnegative, replacing the desired-signal gain by its sampled minimum and each interference gain by its sampled maximum yields sufficient conditions for all constraints in \eqref{eq:scenario_sinr_constraints}. 
This formulation may be more conservative than the finite-scenario LP because the individual extrema need not occur at the same angle. The resulting SINR guarantee applies only to the sampled angles and the assumed range; it does not extend to angles between samples, posterior tails, or range estimation errors. Consequently, the Gaussian angular coverage probability alone does not determine the probability of satisfying the SINR requirement.
Define the illumination coefficients as \(\eta_{l,d}=P_tN_tN_rG_{N_t}(\Theta_d,\widehat\theta_l)\) and \(\eta_d^s=P_tN_tN_r\). For a fixed sensing direction and C4S/S4S mode, the optimization problem is given by
\begin{subequations}\label{eq:power_lp}
\begin{align}
\max_{\boldsymbol\rho_d^c,\rho_d^s}\quad
&\sum_{l=1}^{P}\eta_{l,d}\rho^c_{l,d}
+\chi_d\eta_d^s\rho_d^s\\
\mathrm{s.t.}\quad
&\underline a_{p,p}\rho^c_{p,d}\geq\Gamma_{\min}
\left(\sum_{l\ne p}\overline a_{p,l}\rho^c_{l,d}\right.\nonumber\\
&\hspace{21mm}\left.+\chi_d\overline a_{p,d}^{s}\rho_d^s+\sigma_c^2\right),
\quad\forall p,
\label{eq:power_lp_sinr}\\
&\sum_{p=1}^{P}\rho^c_{p,d}+\chi_d\rho_d^s\leq1,\\
&\boldsymbol\rho_d^c\succeq\mathbf0,\\
&0\leq\rho_d^s\leq\chi_d\overline\rho_d^s.
\end{align}
\end{subequations}
Visual confidence influences sensing-direction selection through \eqref{eq:scan_utility}, whereas the objective in \eqref{eq:power_lp} quantifies illumination for a fixed direction. Since the objective and constraints are affine in the power allocation variables, \eqref{eq:power_lp} is a linear program \cite{boyd2004}. The solution procedure derived in \textbf{Appendix~\ref{app:power_lp}} involves a spectral-radius test, the inversion of a \(P\times P\) matrix, and the evaluation of \(P\) communication-slack directions and one S4S direction in the power allocation space. The resulting computational complexity is \(O(P^3)\) per sensing direction. When the problem is infeasible, all evaluated QoS methods employ the same equal-share fallback policy. Allocation failures and interference-risk constraint violations are recorded, and no sampled-scenario QoS guarantee is claimed under the fallback policy.
\section{Environment-Aware Receiver Processing}



As illustrated in Fig.~\ref{fig:overall_algorithm_flow}, the receiver-side processing operates on the echo snapshots generated by the transmit decisions in Section~IV. The resulting RF detections are
then fed back to update the RF history for the next slot.

\subsection{Narrowband Cluttered Echo}

The unit-norm receive combining vector for direction \(\Theta_d\) is defined as \(\mathbf w_d^r=\mathbf a_{N_r}(\Theta_d)/\sqrt{N_r}\). The combined noise is therefore \(\widetilde z_{d,q}=(\mathbf w_d^r)^H\mathbf n_{d,q}\sim\mathcal{CN}(0,\sigma_s^2)\). The normalized complex field gains are defined as
\begin{equation}
\begin{aligned}
g_d^r(\theta)&=\frac{\mathbf a_{N_r}^H(\Theta_d)
\mathbf a_{N_r}(\theta)}{N_r},
g_p^c(\theta)=\frac{\mathbf a_{N_t}^H(\theta)
\mathbf a_{N_t}(\widehat\theta_p)}{N_t},\\
g_d^s(\theta)&=\frac{\mathbf a_{N_t}^H(\theta)
\mathbf a_{N_t}(\Theta_d)}{N_t}.
\end{aligned}
\end{equation}
Applying the receive combiner to \eqref{eq:received_array_echo} and substituting \eqref{eq:narrowband_transmit_signal} yield the explicit echo representation in \eqref{eq:full_received_echo}. The resulting four terms correspond to target returns induced by the communication and dedicated sensing signals, followed by their respective clutter returns. Each term retains both the transmit and receive array responses. For subsequent receiver processing, we formulate an equivalent complex-baseband model assuming mutually orthogonal, unit-energy pilot sequences and negligible Doppler-induced phase variation within each pilot sequence. The equivalent response energy \(E_d(\theta)\) and scalar response \(\phi_d(\theta)\) associated with each reflector are defined as follows:
\begin{figure*}[!b]
\noindent\rule{\textwidth}{0.4pt}
\vspace{-0.5em}

\begingroup
\small
\begin{equation}
\begin{aligned}
\widetilde y_{d,q}
&=(\mathbf w_d^r)^H\mathbf y_{d,q}^{r}
=\sum_{k=1}^{K_d}\alpha_k e^{j2\pi f_{D,k}qT_p}
\sqrt{N_r}g_d^r(\theta_k)\mathbf a_{N_t}^H(\theta_k)\mathbf x_{d,q}
+\sum_{i=1}^{L_c}\beta_i\sqrt{N_r}g_d^r(\theta_i^{\rm cl})
\mathbf a_{N_t}^H(\theta_i^{\rm cl})\mathbf x_{d,q}+\widetilde z_{d,q}\\
&=\sum_{k=1}^{K_d}\sum_{p=1}^{P}\alpha_k e^{j2\pi f_{D,k}qT_p}
\sqrt{\rho_{p,d}^c P_tN_tN_r}
\times g_d^r(\theta_k)g_p^c(\theta_k)s_{p,d,q}^c
+\chi_d\sum_{k=1}^{K_d}\alpha_k e^{j2\pi f_{D,k}qT_p}
\sqrt{\rho_d^sP_tN_tN_r}
\times g_d^r(\theta_k)g_d^s(\theta_k)s_{d,q}^s\\
&+\sum_{i=1}^{L_c}\sum_{p=1}^{P}\beta_i\sqrt{\rho_{p,d}^cP_tN_tN_r}
\times g_d^r(\theta_i^{\rm cl})g_p^c(\theta_i^{\rm cl})s_{p,d,q}^c
+\chi_d\sum_{i=1}^{L_c}\beta_i\sqrt{\rho_d^sP_tN_tN_r}
\times g_d^r(\theta_i^{\rm cl})g_d^s(\theta_i^{\rm cl})s_{d,q}^s
+\widetilde z_{d,q}.
\end{aligned}
\label{eq:full_received_echo}
\end{equation}
\endgroup

\end{figure*}
\begin{subequations}
\begin{equation}
\begin{aligned}
E_d(\theta)={}&P_tN_tN_r|g_d^r(\theta)|^2\\
&\times\left[\sum_{p=1}^{P}\rho_{p,d}^c|g_p^c(\theta)|^2
+\chi_d\rho_d^s|g_d^s(\theta)|^2\right].
\end{aligned}
\label{eq:echo_illumination_power}
\end{equation}
\begin{equation}
\phi_d(\theta)=\sqrt{E_d(\theta)}\,e^{j\psi_d(\theta)}.
\label{eq:compact_angular_response}
\end{equation}
\end{subequations}
The phase reference is chosen as \(\psi_d(\theta)=\arg g_d^r(\theta)\), with an arbitrary value when \(E_d(\theta)=0\). By construction, the equivalent scalar response satisfies \(|\phi_d(\theta)|^2=E_d(\theta)\) and is used consistently in the numerical receiver. This response provides an energy-preserving processing model for each reflector, rather than a coherent combination of the individual waveform branches. Accordingly, the clutter estimator and numerical evaluation employ the equivalent complex-baseband observation model
\begin{align}
y_{d,q}={}&\sum_{k=1}^{K_d}\alpha_k\phi_d(\theta_k)
e^{j2\pi f_{D,k}qT_p}
+\sum_{i=1}^{L_c}\beta_i\phi_d(\theta_i^{\rm cl})+z_{d,q},
\label{eq:cluttered_echo}
\end{align}
where $z_{d,q}\sim\CN(0,\sigma_s^2)$ is the adopted effective snapshot noise,
independent across directions and snapshots. 
Equation~\eqref{eq:cluttered_echo} preserves the per-reflector response energy but is not
an exact coherent reduction of the waveform-resolved branches in \eqref{eq:full_received_echo},
since summed output powers do not determine a complex Gaussian observation.

\subsection{RF-Calibrated Semantic Clutter Reconstruction}

Let \(\mathcal I_c=\{i_1,\ldots,i_{L_v}\}\) denote the ordered set of screened clutter candidates, where \(L_v=|\mathcal I_c|\). Stacking the \(M=Q_nQ_s\) scalar observations in \eqref{eq:cluttered_echo} yields
\begin{equation}
\mathbf y=\boldsymbol\Phi_v\boldsymbol\beta+\mathbf u_{\rm dyn}+\mathbf z,
\label{eq:stacked_equivalent_model}
\end{equation}
where $\mathbf y,\mathbf u_{\rm dyn},\mathbf z\in\mathbb C^M$,
$\boldsymbol\Phi_v\in\mathbb C^{M\times L_v}$, and
$\boldsymbol\beta=(\beta_{i_1},\ldots,\beta_{i_{L_v}})^T\in\mathbb C^{L_v}$
contains the coefficients on the visual support. Column $j$ repeats
$\phi_d(\widehat\theta_{i_j}^v)$ over the snapshots of each selected direction.
Vision supplies support and classes, while RF observations estimate these coefficients.
The screened support need not coincide with the \(L_c\) physical clutter cells. The effects of omitted clutter components and target-associated columns incorrectly classified as static are analyzed in \textbf{Appendix~\ref{app:map_estimator}}. Let \(\mathbf R_z\in\mathbb C^{M\times M}\) denote the noise covariance, with \(\mathbf R_z=\sigma_s^2\mathbf I_M\) under the adopted white-noise model. The class-conditioned prior precision matrix is
$\boldsymbol\Lambda=\diag(\nu_{c_{i_1}}^{-1},\ldots,\nu_{c_{i_{L_v}}}^{-1})
\in\mathbb R^{L_v\times L_v}$.
Under a nominal clutter-only Gaussian likelihood and the proper complex Gaussian prior \(\boldsymbol\beta\sim\mathcal{CN}(\mathbf0,\boldsymbol\Lambda^{-1})\) \cite{goodman1963,kay1993}, the maximum a posteriori estimate is obtained as
\begin{equation}
\widehat{\boldsymbol\beta}=
\arg\min_{\boldsymbol\beta}
\|\mathbf y-\boldsymbol\Phi_v\boldsymbol\beta\|_{\mathbf R_z^{-1}}^2
+\boldsymbol\beta^H\boldsymbol\Lambda\boldsymbol\beta,
\end{equation}
which admits the closed-form solution
\begin{equation}
\widehat{\boldsymbol\beta}=
\left(\boldsymbol\Phi_v^H\mathbf R_z^{-1}\boldsymbol\Phi_v
+\boldsymbol\Lambda\right)^{-1}
\boldsymbol\Phi_v^H\mathbf R_z^{-1}\mathbf y.
\label{eq:clutter_map_solution}
\end{equation}
The derivation is provided in \textbf{Appendix~\ref{app:map_estimator}}. In the numerical implementation, a small diagonal loading term is added to the matrix being inverted to improve numerical stability. The residual after semantic clutter cancellation is
\begin{equation}
\mathbf Y^{\rm sem}=\mathbf Y-\operatorname{unstack}
(\boldsymbol\Phi_v\widehat{\boldsymbol\beta})
\label{eq:semantic_residual}
\end{equation}
where \(\operatorname{unstack}(\cdot)\) restores the original arrangement of sensing directions and snapshots.

\subsection{Motion-Aware Residual Static Filtering}

To protect visually identified moving objects during residual clutter suppression, we define the protection weight for each selected sensing direction \(d\) as
\begin{equation}
m_d=1-\prod_{j\in\mathcal V}
\left[1-\gamma_j^vp_j^{\rm mov}
e^{-\frac{(\Theta_d-\widehat\theta_j^v)^2}{2\widetilde\sigma_j^2}}\right],
\label{eq:45}
\end{equation}
where $\widetilde\sigma_j=\max(\sigma_{\theta,j}^v,\sigma_{\rm support})$ and $\sigma_{\rm support}$ is the minimum angular width of the motion-protection profile. We estimate the residual static component as
\begin{equation}
\widehat c_d^{\rm res}=(1-m_d)\frac{1}{Q_s}
\sum_{q=0}^{Q_s-1}Y^{\rm sem}_{d,q},
\label{eq:motion_aware_static_estimate}
\end{equation}
and the filtered observation is obtained as
\begin{equation}
Y^{\rm clean}_{d,q}=Y^{\rm sem}_{d,q}-\widehat c_d^{\rm res}.
\label{eq:clean_echo}
\end{equation}
The cases \(m_d=0\) and \(m_d=1\) correspond to full temporal mean subtraction and no residual subtraction, respectively. This weighting also protects visually identified moving vehicles with zero radial Doppler. For a target component \(Ae^{jq\omega_D}\), the normalized magnitude of its temporal mean is
\begin{equation}
A_{\rm avg}(\omega_D)=
\left|\frac{\sin(Q_s\omega_D/2)}{Q_s\sin(\omega_D/2)}\right|,
\qquad \omega_D=2\pi\frac{2v}{\lambda}T_p,
\label{eq:temporal_average_response}
\end{equation}
with the ratio defined by continuity when the denominator vanishes. The fraction of burst-averaged target power retained after filtering is
\begin{equation}
\eta_{\rm MA}(\omega_D,m_d)=
1-(1-m_d^2)A_{\rm avg}^2(\omega_D).
\label{eq:motion_aware_retention}
\end{equation}
At $v=0$, $\omega_D=0$ and $A_{\rm avg}=1$, so $\eta_{\rm MA}=m_d^2$ independently of $Q_s$ for a fixed protection weight. For a fixed nonzero Doppler that does not alias to zero, increasing $Q_s$ reduces the envelope of the temporal-mean response and generally improves target retention. The Dirichlet-kernel lobes and nulls prevent strict monotonicity; longer bursts better separate nonzero-Doppler returns from static components, while zero Doppler remains difficult. The protection weight in \eqref{eq:45} jointly incorporates visual motion probability, detection confidence, and angular proximity. \textbf{Appendix~\ref{app:slow_target}} derives this filter-level power-retention result and its low-Doppler expansion. The effects of the preceding MAP clutter cancellation and the resulting detection performance are evaluated separately.














\begin{algorithm}[t]
\caption{Per-Slot Vision--RF Environment-Aware ISAC}
\label{alg:vision_rf_isac}
\begin{algorithmic}[1]
\REQUIRE Prior states, visual observations, RF history,
semantic map, and scan ages
\ENSURE Communication schedule and detected targets
$(\hat{\theta}_k,\hat{v}_k)$
\FOR{each communication user $p$}
    \STATE Predict by (6) and fuse available observations by (9)--(10)
\ENDFOR
\STATE Estimate $r_v$ and determine $Q_n,Q_{\rm exp}$ by (21) and (23)
\STATE Select sensing directions using visual/RF priorities and
oldest-first exploration
\FOR{each selected direction $d$}
    \STATE Determine C4S/S4S modes by (28)--(29), solve (36),
    and select the feasible mode
    \STATE Transmit (11) and collect $Q_s$ matched-filtered snapshots
\ENDFOR
\STATE Estimate and subtract semantic clutter by (42)--(44)
\STATE Apply motion-aware residual filtering by (45)--(47)
\STATE Form the Angle--Doppler map by (50) and detect targets by (51)--(52)
\STATE Update RF history and scan ages for the next slot
\end{algorithmic}
\end{algorithm}
\subsection{Angle--Doppler Processing and 2-D CFAR}

The BS applies the standard slow-time Doppler discrete Fourier transform (DFT) \cite{richards2014} to the clean direction--snapshot matrix:
\begin{equation}
H_{\rm AD}[d,b]=\frac{1}{\sqrt{Q_s}}
\sum_{q=0}^{Q_s-1}Y^{\rm clean}_{d,q}
e^{-j2\pi bq/Q_s}.
\label{eq:angle_doppler_transform}
\end{equation}
After fast Fourier transform (FFT) shifting, each Doppler bin is associated with a frequency \(f_b\) and the corresponding radial velocity \(v_b=\lambda f_b/2\). Two-dimensional cell-averaging CFAR (CA-CFAR) detection \cite{gandhi1988} is performed on \(|H_{\mathrm{AD}}[d,b]|^2\). To accommodate nonuniform sensing directions, the angular training and guard regions are defined using physical angular separations. For the cell under test (CUT) \((d,b)\), the training set is
\begin{equation}
\begin{aligned}
&\mathcal T_{d,b}=\{(d',b'):\;|\Theta_{d'}-\Theta_d|\leq\Delta_{\rm tr}^\theta, |b'-b|_{Q_s}\leq M_{\rm tr}^b,\\
&\quad\neg(|\Theta_{d'}-\Theta_d|\leq\Delta_{\rm g}^\theta,\,
|b'-b|_{Q_s}\leq M_{\rm g}^b)\},
\end{aligned}
\label{eq:physical_angle_training_set}
\end{equation}
Here $\Delta_{\rm tr}^\theta$ and $\Delta_{\rm g}^\theta$ are the angular training and guard half-widths; $M_{\rm tr}^b$ and $M_{\rm g}^b$ are their Doppler-bin counterparts. The joint guard region is excluded, and \(|b'-b|_{Q_s}\) denotes the circular distance between Doppler bins. Let \(N_{\mathrm{tr}}(d,b)=|\mathcal T_{d,b}|\), and let \(\widehat\sigma_{d,b}^2\) denote the average power over the training cells. For a nominal false-alarm probability \(P_{\mathrm{FA}}^{(0)}\), the CA-CFAR scaling factor and detection threshold are \cite{gandhi1988}
\begin{equation}
\begin{aligned}
\alpha_{d,b}&=N_{\rm tr}(d,b)
\left[(P_{\rm FA}^{(0)})^{-1/N_{\rm tr}(d,b)}-1\right],
\\
\tau_{d,b}&=\alpha_{d,b}\widehat\sigma^2_{d,b}.
\end{aligned}
\label{eq:physical_angle_cfar_threshold}
\end{equation}
This scaling follows from the standard assumption that the CUT and training-cell powers are independent and exponentially distributed with a common mean \cite{gandhi1988}. Local maxima exceeding their corresponding thresholds yield the target estimates \((\widehat\theta_k,\widehat v_k)\). The actual training-cell count is used at angular boundaries and in sparsely sampled regions; CUTs with fewer than the prescribed minimum number of training cells are excluded. Correlation and clutter nonhomogeneity may cause the achieved false-alarm probability to differ from its nominal value \cite{gandhi1988}.
\textbf{Algorithm~\ref{alg:vision_rf_isac}} summarizes the estimation, transmission, and detection procedures executed in each slot. The resulting detections are fed back to update the RF history, scan-age information, and target priors for the next slot, thereby coupling the transmit-side scheduling in Section IV with the environment-aware receiver processing in Section V.

\begin{figure*}[t]
\centering
\includegraphics[width=0.8\textwidth]{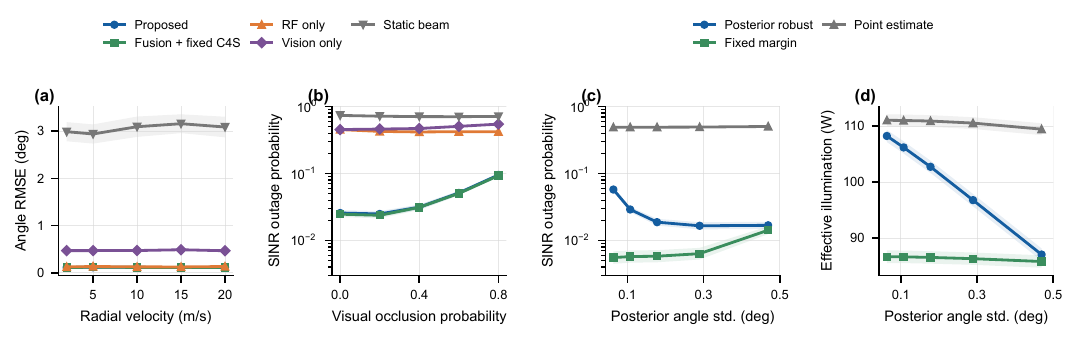}
\caption{Dynamic-user communication performance: (a) angle RMSE; (b) outage versus visual occlusion; (c) posterior-robust outage; (d) effective sensing illumination.}
\label{fig:communication_results}
\end{figure*}


\section{Numerical Results}

\textbf{Table~\ref{tab:simulation_parameters}} lists the simulation parameters. The SBS visual/RF weights and maximum exploration reserve were selected from eight fixed candidates on separate calibration scenes, using detection probability averaged across scan budgets. Candidates had to limit degradation relative to the original SBS to 0.03 in both low-occlusion and sparse-target settings. The selected parameters were frozen before evaluation, without per-scenario tuning. The codebook size, budget bounds, coverage threshold, and remaining constants are fixed implementation parameters, rather than theoretically optimal values. Five settings evaluate
dynamic-user communication, adaptive scanning, clutter suppression,
and slow-target detection using identical scenes and observations
across methods. Continuous metrics are reported as means with 95\%
confidence intervals, and detection and outage probabilities with
Wilson intervals \cite{wilson1927}. Comparisons involving the proposed receiver
and revised SBS use paired bootstrap intervals over complete episodes
\cite{efron1993}.
\begin{table}[t]
\centering
\caption{Shared simulation parameters.}
\label{tab:simulation_parameters}
\footnotesize
\setlength{\tabcolsep}{3.2pt}
\begin{tabular}{@{}p{0.58\columnwidth}p{0.36\columnwidth}@{}}
\toprule
Parameter & Value \\
\midrule
Carrier frequency; transmit power & $28$~GHz; $1$~W \\
Transmit/receive ULA size & $16/16$ elements \\
Element spacing & $\lambda/2$ \\
Communication/sensing noise power & $10^{-9}$~W \\
Slot duration; snapshot interval & $0.1$~s; $50~\mu$s \\
Snapshots per sensing direction & $64$ \\
Angular field of view & $[-60^\circ,60^\circ]$ \\
Codebook size $Q_{\rm full}$ & $121$ \\
Scan budgets $Q_{\min},Q_{\max}$ & $8,24$ \\
Exploration $Q_{\rm exp,min},Q_{\rm exp,max}$ & $4,6$ \\
Exploration-budget slope $\kappa_{\rm exp}$ & $2$ \\
Coverage/uncertainty $\tau_{\rm cov},\kappa_Q$ & $0.9,1$ \\
Weights $(\lambda_{v0},\lambda_{r0},\lambda_{u0},\lambda_{e0})$ & $(4,2,0.2,0.15)$ \\
Exploration-weight slope $\kappa_e$ & $0.35$ \\
Widths $\sigma_{\rm ref},\sigma_h$; decay $\delta$ & $3^\circ,3^\circ$; $0.9$ \\
Visible-field endfire margin $\delta_\theta$ & $10^{-9}$~rad \\
Motion-profile width $\sigma_{\rm support}$ & $4^\circ$ \\
Static/motion thresholds $\tau_{\rm static},\tau_{\rm mov}$ & $0.70,0.30$ \\
Moving-ROI margin $\Delta_{\rm mask}$ & $4^\circ$ \\
Vision/RF angle standard deviation & $1.0^\circ/0.6^\circ$ \\
Vision/RF detection probability & $0.90/0.95$; vision $1$ in Exp.~4 \\
Robust QoS multiplier/samples & $\kappa_c=2$; $S_c=5$ \\
Fixed QoS angular margin/samples & $1^\circ$; $5$ \\
2-D CA-CFAR design $P_{\rm FA}$ & $10^{-3}$ \\
Communication episodes & $100$ (Exp.~1, 6) \\
Scanning/clutter/slow-target episodes & $300/200/200$ \\
\bottomrule
\end{tabular}
\end{table}

Fig.~\ref{fig:communication_results}(a)--(b) compares the performance of static beams, vision-only tracking, RF-only tracking, and the proposed vision--RF fused tracking. As shown in Fig.~\ref{fig:communication_results}(a), at a vehicle speed of 20 m/s, the proposed scheme achieves an angular root-mean-square error (RMSE) of $0.114^\circ$, compared with $3.087^\circ$ for static beams and $0.465^\circ$ for vision-only tracking. This is because motion prediction captures vehicle dynamics, while RF observations compensate for unreliable visual measurements. As the visual occlusion rate increases, the outage probability of vision-only tracking increases significantly, whereas the proposed scheme remains robust and reduces the outage probability from 0.5475 to 0.0953 at an occlusion rate of 0.8. It is also observed that dynamic C4S/S4S selection achieves an outage probability comparable to fixed selection, while reducing the C4S ratio from $42.62\%$ to $4.36\%$ and increasing the effective sensing illumination from 105.27 to 106.36 W. This indicates that the proposed dynamic strategy improves sensing performance at the cost of only a moderate communication SINR reduction.
Fig.~\ref{fig:communication_results}(c)--(d) compares point control, fixed-$1^\circ$-margin control, and posterior-robust control under identical states and realizations. As shown in Fig.~\ref{fig:communication_results}(c), at a posterior angular standard deviation of $0.0633^\circ$, posterior-robust control reduces the outage probability from 0.4903 to 0.0576 relative to point control, while achieving 108.22 W sensing illumination. Compared with the fixed-margin scheme, it provides stronger sensing illumination with a slightly higher outage probability. This is because the posterior interval adapts to the actual angular uncertainty, avoiding unnecessarily conservative protection. As shown in Fig.~\ref{fig:communication_results}(d), increasing \(\kappa_c\) from 0.5 to 3 reduces the outage probability from 0.2902 to 0.0062, while the sensing illumination decreases from 110.68 to 104.57 W. This is because a larger uncertainty interval requires more communication power to satisfy the robust SINR constraints. Therefore, increasing \(\kappa_c\) improves communication robustness at the cost of reduced sensing illumination.

\begin{figure*}[tb]
\centering
\includegraphics[width=0.7\textwidth]{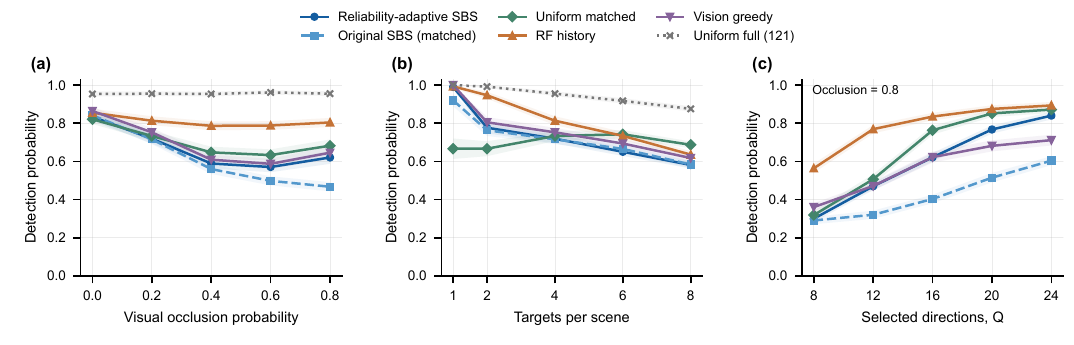}
\caption{SBS detection versus (a) occlusion, (b) target density, and (c) budget at $0.8$ occlusion. Budgets are paired except full scanning ($121$ directions). Shading: 95\% episode-level intervals; $300$ episodes/point.}
\label{fig:adaptive_sbs}
\end{figure*}

\begin{figure*}[t]
\centering
\includegraphics[width=0.8\textwidth]{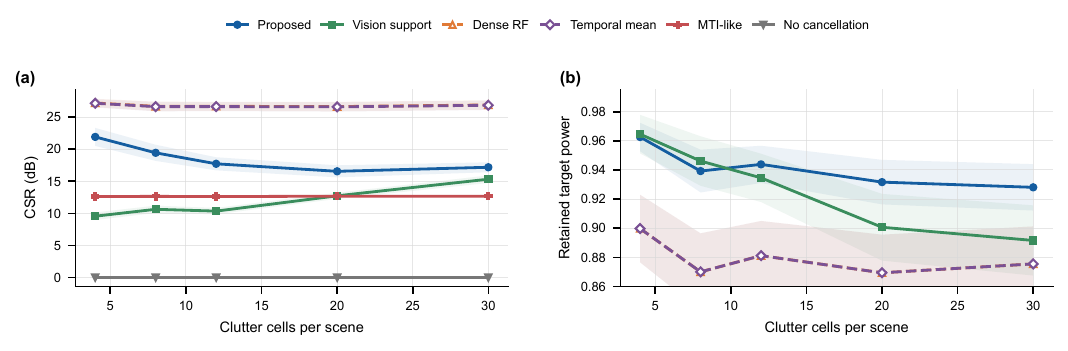}
\caption{Clutter suppression and robustness: (a) CSR and (b) retained target power versus clutter density, using the original results.}
\label{fig:clutter_suppression}
\end{figure*}

Fig.~\ref{fig:adaptive_sbs} shows the performance comparison of different SBS schemes under visual occlusion, varying target densities, and scan budgets. For a fair comparison, all budget-constrained schemes select the same number of sensing directions, while full scanning visits all 121 directions. As shown in Fig.~\ref{fig:adaptive_sbs}, the detection probability generally decreases as visual occlusion becomes more severe. At an occlusion rate of 0.8, the proposed SBS improves the detection probability from 0.468 to 0.621 compared with the original SBS. This is because more weight is assigned to RF history when visual observations become unreliable. Meanwhile, the proposed SBS selects only 17.18 directions on average, reducing the sensing time from 0.3872 to 0.0550 s and the scan energy from 0.1936 to 0.0275 J compared with full scanning. It is also observed that removing RF history and using fixed weights degrade the detection probability to 0.458 and 0.601, respectively. These results demonstrate that the proposed reliability-adaptive SBS effectively improves robustness to visual blockage while significantly reducing sensing overhead.

Fig.~\ref{fig:clutter_suppression} shows the comparison of clutter suppression and target retention for different receiver schemes. As shown in Fig. 6, with 30 clutter cells, the proposed receiver achieves a clutter suppression ratio (CSR) of 17.19 dB and retains 0.928 of the target power, compared with 26.86 dB and 0.876 for temporal averaging. This is because the motion-aware weighting reduces temporal subtraction in regions likely to contain moving vehicles, thereby preserving more slow-target energy. However, the weaker clutter suppression leaves more residual clutter, resulting in a lower detection probability of 0.795 compared with 0.925 for temporal averaging. This tradeoff arises because motion-aware weighting deliberately reduces temporal subtraction in regions likely to contain moving vehicles. It preserves target energy but also leaves a larger residual static component in those protected regions, which can limit detection despite improved retention.


\begin{figure*}[tb]
\centering
\includegraphics[width=0.8\textwidth]{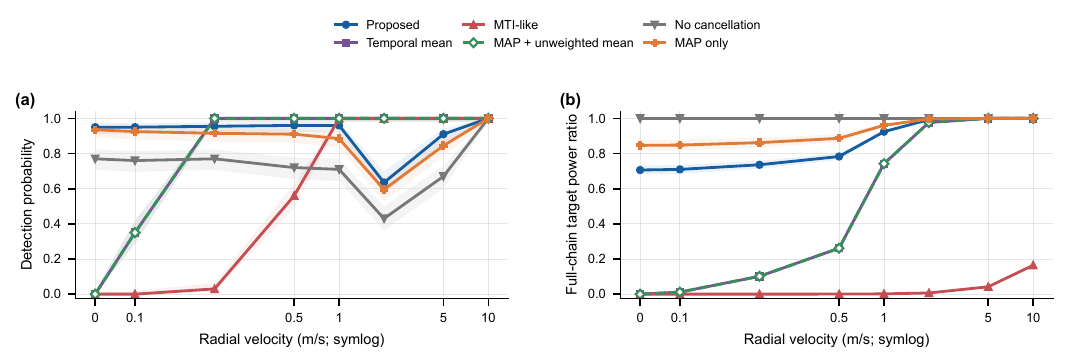}
\caption{Slow-target protection at $p^{\rm mov}=1$: (a) detection probability; (b) full-chain retained target power. Velocity axes use a symmetric logarithmic scale.}
\label{fig:slow_target}
\end{figure*}

Fig.~\ref{fig:slow_target} compares the performance of six receiver schemes with an available visual vehicle prior (\(p^{\mathrm{mov}}=1\)). As shown in Fig.~\ref{fig:slow_target}, the detection performance of conventional temporal cancellation degrades significantly as the radial velocity approaches zero, since slow-moving target components are increasingly removed together with static clutter. In contrast, the proposed motion-aware receiver maintains effective target preservation in the low-Doppler region. At zero velocity, it retains 0.706 of the target power and achieves a detection probability of 0.950, while unweighted temporal-mean and MTI-like processing fail to detect the target. It is also observed that MAP-only processing retains more target power, reaching 0.846, and achieves a comparable detection probability of 0.935. This indicates that the motion-aware stage mainly improves robustness to severe low-Doppler attenuation after clutter reconstruction, rather than providing a uniform gain over MAP-only processing.

\section{Conclusion}


A vision--RF framework is proposed for dynamic-user communication and target sensing in roadside clutter under a narrowband ISAC model. We have developed covariance-aware state fusion to guide beam steering, C4S/S4S mode selection, and power allocation under sampled-angle QoS constraints. To reduce sensing overhead, we have designed a reliability-adaptive scanning scheme that combines visual information and RF history with reserved exploration within a bounded scan budget. At the receiver, clutter coefficients are estimated over visually identified support, followed by motion-aware residual filtering before angle--Doppler CFAR detection. Our numerical results show improved tracking over static and vision-only beam control, reduced scanning cost relative to full scanning, and reduced slow-target attenuation relative to temporal cancellation when visual priors are available. 
Reliability-adaptive scanning mitigates blockage degradation relative
to fixed-weight scanning, while revealing tradeoffs between exploration
and immediate detection and between clutter suppression and target
preservation. Future work will investigate measured roadside data and
visual--RF association errors.

\appendices

\section{Derivation of the Slow-Target Retention Ratio}
\label{app:slow_target}

For a target-only burst $y_q=Ae^{jq\omega_D}$ with $A\ne0$ and integer
$Q_s\geq1$, the finite geometric sum \cite{richards2014} gives
\begin{equation}
D_{Q_s}(\omega_D)=\frac{1}{Q_s}\sum_{q=0}^{Q_s-1}e^{jq\omega_D}
=\frac{e^{j(Q_s-1)\omega_D/2}\sin(Q_s\omega_D/2)}
{Q_s\sin(\omega_D/2)}.
\label{eq:dirichlet_kernel}
\end{equation}
The finite sum defines the removable singularities, including $D_{Q_s}(0)=1$,
and its magnitude is \eqref{eq:temporal_average_response}.
For fixed $m_d\in[0,1]$, decompose the burst into its temporal mean
$\bar y=AD_{Q_s}$ and the orthogonal zero-mean component. The residual filter
$\widetilde y_q=y_q-(1-m_d)\bar y$ scales only the mean by $m_d$, so
\begin{equation}
\eta_{\rm MA}(\omega_D,m_d)
=\frac{\sum_q|\widetilde y_q|^2}{\sum_q|y_q|^2}
=1-(1-m_d^2)|D_{Q_s}(\omega_D)|^2.
\label{eq:appendix_motion_aware_retention}
\end{equation}
Since $|D_{Q_s}|\leq1$, $m_d^2\leq\eta_{\rm MA}\leq1$; $m_d=0$ gives
ordinary mean subtraction and $m_d=1$ gives unit retention. In particular,
\begin{equation}
\lim_{\omega_D\rightarrow0}\eta_{\rm MA}(\omega_D,m_d)=m_d^2.
\label{eq:motion_aware_zero_doppler}
\end{equation}
Expanding $|D_{Q_s}(\omega_D)|^2$ at zero for fixed $Q_s$ yields
\begin{equation}
\eta_{\rm MA}=m_d^2+(1-m_d^2)
\frac{Q_s^2-1}{12}\omega_D^2+\mathcal O(\omega_D^4).
\label{eq:motion_aware_taylor}
\end{equation}
The quadratic approximation requires $Q_s|\omega_D|\ll1$ when comparing
burst lengths. At a common Doppler, the ROI power ratio is the
$|A_d|^2$-weighted mean of the per-direction ratios and tends to the
corresponding weighted mean of $m_d^2$ at zero Doppler. This target-only
filter result excludes MAP cancellation, CFAR detection, and RF
identification of zero-Doppler motion.

\section{Derivation of the MAP Clutter Estimator}
\label{app:map_estimator}

Condition on the screened dictionary and fixed prior/noise parameters in
\eqref{eq:stacked_equivalent_model}. The nominal clutter-only model
\cite{goodman1963,kay1993} is
\begin{equation}
\begin{aligned}
\mathbf y\mid\boldsymbol\beta&\sim\CN(\boldsymbol\Phi_v\boldsymbol\beta,\mathbf R_z),\\
\boldsymbol\beta&\sim\CN(\mathbf0,\boldsymbol\Lambda^{-1}),
\quad\mathbf R_z\succ0,\quad\boldsymbol\Lambda\succ0,
\end{aligned}
\label{eq:nominal_map_model}
\end{equation}
with independent proper complex Gaussian noise and prior coefficients.
Here $\boldsymbol\Phi_v\in\mathbb C^{M\times L_v}$ uses the ordered screened
support; $\mathbf R_z$ and $\boldsymbol\Lambda$ have dimensions $M\times M$
and $L_v\times L_v$, respectively. Define
\begin{equation}
\mathbf A=\boldsymbol\Phi_v^H\mathbf R_z^{-1}\boldsymbol\Phi_v
+\boldsymbol\Lambda.
\label{eq:map_normal_matrix}
\end{equation}
Minimizing the negative log posterior by complex differentiation
\cite{hjorungnes2007} gives
\begin{equation}
\widehat{\boldsymbol\beta}
=\mathbf A^{-1}\boldsymbol\Phi_v^H\mathbf R_z^{-1}\mathbf y.
\label{eq:appendix_map_unique_solution}
\end{equation}
Since $\mathbf R_z\succ0$ and $\boldsymbol\Lambda\succ0$, $\mathbf A\succ0$;
hence the MAP objective is strictly convex and its stationary point is
unique, even for rank-deficient $\boldsymbol\Phi_v$. The nominal posterior
is $\CN(\widehat{\boldsymbol\beta},\mathbf A^{-1})$. Prior-variance flooring
and diagonal loading enter through the effective precision
$\boldsymbol\Lambda$.

For the actual conditional input
$\mathbf y=\boldsymbol\Phi_v\boldsymbol\beta+\mathbf u_{\rm dyn}+\mathbf z$,
substitution into \eqref{eq:appendix_map_unique_solution} yields
\begin{equation}
\begin{aligned}
\widehat{\boldsymbol\beta}-\boldsymbol\beta
={}&-\mathbf A^{-1}\boldsymbol\Lambda\boldsymbol\beta
+\mathbf A^{-1}\boldsymbol\Phi_v^H\mathbf R_z^{-1}\mathbf u_{\rm dyn}\\
&+\mathbf A^{-1}\boldsymbol\Phi_v^H\mathbf R_z^{-1}\mathbf z,
\end{aligned}
\label{eq:map_error_decomposition}
\end{equation}
separating prior shrinkage, target leakage, and noise error. With
$\mathbf K=\mathbf A^{-1}\boldsymbol\Phi_v^H\mathbf R_z^{-1}$,
submultiplicativity gives
\begin{equation}
\|\Delta\boldsymbol\beta_{\rm tar}\|_2
\leq\|\mathbf A^{-1}\boldsymbol\Phi_v^H\mathbf R_z^{-1}\|_2
\|\mathbf u_{\rm dyn}\|_2.
\label{eq:target_leakage_bound}
\end{equation}
The target-only residual is $(\mathbf I_M-\boldsymbol\Phi_v\mathbf K)
\mathbf u_{\rm dyn}$. Averaging over noise leaves both deterministic
error terms; the sampling covariance $\mathbf K\mathbf R_z\mathbf K^H$
differs from the nominal posterior covariance $\mathbf A^{-1}$.
Neither covariance includes squared deterministic bias or unmodeled-target
uncertainty.

An omitted clutter term $\Delta\boldsymbol\Phi\boldsymbol\beta_{\rm miss}$
enters the coefficient error through $\mathbf K$ and also remains in the
residual. A false-static target column increases dictionary--target
coherence and can increase leakage. These support errors lie outside the
nominal posterior covariance; \eqref{eq:clutter_eligibility} reduces their
occurrence but does not eliminate them.

\section{Feasibility Analysis of the Per-Direction LP}
\label{app:power_lp}

For fixed direction, mode, and envelope coefficients, normalize
\eqref{eq:power_lp_sinr} by $\underline a_{p,p}>0$:
\begin{equation}
\begin{aligned}
F_{p,l}&=\begin{cases}
\Gamma_{\min}\overline a_{p,l}/\underline a_{p,p},&l\ne p,\\
0,&l=p,
\end{cases}\\
u_p&=\frac{\Gamma_{\min}\sigma_c^2}{\underline a_{p,p}},\qquad
v_p=\frac{\Gamma_{\min}\overline a_{p,d}^{s}}{\underline a_{p,p}}.
\end{aligned}
\label{eq:normalized_power_coefficients}
\end{equation}
The coupled SINR constraints are
\begin{equation}
(\mathbf I-\mathbf F)\boldsymbol\rho_d^c
\succeq\mathbf u+\chi_d\mathbf v\rho_d^s.
\label{eq:appendix_power_inequality}
\end{equation}
Here $\mathbf F\succeq\mathbf0$, $\mathbf v\succeq\mathbf0$, and positive
SINR thresholds, noise, and desired-link gains ensure $\mathbf u\succ\mathbf0$.
For fixed finite $\rho_d^s\geq0$, classical interference-coupled power control
\cite{yates1995} and nonnegative-matrix theory \cite{horn2013} give feasibility
before the total-power constraint if and only if $\varrho(\mathbf F)<1$.
Then $\mathbf M=(\mathbf I-\mathbf F)^{-1}\succeq\mathbf0$ and
\begin{equation}
\boldsymbol\rho_{d,\min}^c(\rho_d^s)
=\mathbf M(\mathbf u+\chi_d\mathbf v\rho_d^s).
\label{eq:minimum_communication_power}
\end{equation}
For C4S, $\chi_d=0$ and $\rho_d^s=0$, so
\begin{equation}
\text{C4S feasible}\ \Longleftrightarrow\
\varrho(\mathbf F)<1,\quad\mathbf1^T\mathbf M\mathbf u\leq1.
\label{eq:c4s_power_feasibility}
\end{equation}
For S4S, $\chi_d=1$. Under the same spectral-radius and unit-budget
conditions, intersecting the budget with the fixed risk cap
$0\leq\overline\rho_d^s\leq1$ gives $0\leq\rho_d^s\leq\rho_{d,\max}^s$, where
\begin{equation}
\rho_{d,\max}^{s}=\min\left\{\overline\rho_d^s,
\frac{1-\mathbf1^T\mathbf M\mathbf u}
{1+\mathbf1^T\mathbf M\mathbf v}\right\}.
\label{eq:maximum_s4s_power}
\end{equation}
The maximum sensing power is $P_t\rho_{d,\max}^s$ in watts. It is strictly
positive precisely when $\varrho(\mathbf F)<1$,
$\mathbf1^T\mathbf M\mathbf u<1$, and $\overline\rho_d^s>0$; an exhausted
budget or zero cap allows only zero sensing. Neither mode is feasible if
the spectral-radius condition fails or $\mathbf1^T\mathbf M\mathbf u>1$.

\emph{Proof sketch:} Any feasible positive communication vector satisfies
$\mathbf F\boldsymbol\rho_d^c\prec\boldsymbol\rho_d^c$, which implies
$\varrho(\mathbf F)<1$ even for reducible $\mathbf F$.
Conversely, premultiplying \eqref{eq:appendix_power_inequality} by the
nonnegative resolvent gives \eqref{eq:minimum_communication_power}.
All feasible vectors have the form
$\boldsymbol\rho_d^c=\mathbf M(\mathbf u+\chi_d\mathbf v\rho_d^s+\mathbf s)$
with $\mathbf s\succeq\mathbf0$; thus the minimum total power is
$\mathbf1^T\mathbf M\mathbf u+\chi_d\rho_d^s(1+\mathbf1^T\mathbf M\mathbf v)$,
yielding the C4S test and S4S cap. After reserving the minimum allocation,
the affine objective is a fractional knapsack over the $P$ communication
slack rays (columns of $\mathbf M$) and the S4S ray
$(\mathbf M\mathbf v,1)$. The implementation orders them by illumination
gain per unit budget and caps the S4S ray at \eqref{eq:maximum_s4s_power}.
This applies to the conservative envelope in \eqref{eq:robust_qos_envelopes},
not a general stacked-scenario LP; the largest feasible sensing fraction
need not maximize illumination.

\bibliographystyle{IEEEtran}
\bibliography{arxiv}

\end{document}